\documentstyle[prl,aps,psfig]{revtex}

\begin{document}
\begin{center}

{\bf Monte Carlo Simulations of the Random-Field Ising Model}

W. C. Barber and D. P. Belanger

Department of Physics, University of California, Santa Cruz, CA 95064, USA

\end{center}

\begin{abstract}

Ising Monte Carlo simulations of
the random-field Ising system
${\rm Fe}_{0.80}{\rm Zn}_{0.20}{\rm F}_2$
are presented for $H=10$~T.
The specific heat critical behavior is consistent with $\alpha \approx 0$
and the staggered magnetization
with $\beta \approx 0.25 \pm 0.03$.

\end{abstract}

\vspace{0.5in}

It has recently been shown experimentally\cite{sbf99} and through Monte
Carlo (MC)
simulations\cite{bb00} that the three dimensional ($d=3$) dilute,
anisotropic antiferromagnet ${\rm Fe}_x{\rm Zn}_{1-x}{\rm F}_2$ in an applied
uniform field exhibits the equilibrium critical behavior of the
random-field Ising model (RFIM)\cite{b98} only for magnetic
concentrations $x > 0.75$.  Although there is some agreement
between previous MC and equilibrium experimental data,
particularly the staggered susceptibility critical exponent, where
$\gamma = 1.7\pm 0.2$ for MC\cite{r95} and $\gamma = 1.58\pm 0.08$ for
experiment\cite{sbf99}, and the correlation length critical exponent,
where $\nu = 1.1\pm 0.2$ for MC\cite{r95} and $\nu= 0.88\pm 0.05$ for
experiment\cite{sbf99}, the experimental value\cite{sb98} of the
specific heat exponent, $\alpha = 0.00 \pm 0.02$, disagrees substantially
with the simulation result\cite{r95} $\alpha = -0.5 \pm 0.2$.  The order
parameter exponent, $\beta =0.00\pm 0.05$ from simulations\cite{r95},
has yet to be measured experimentally.  The MC exponents
violate the Rushbrooke scaling inequality $2\beta +\gamma + \alpha \ge 2$,
indicating possible inconsistencies.  The previous MC studies employed finite
scaling on ferromagnetic lattices of sizes $L \times L \times L$,
with $L \le 16$.  Simulations were made over many random-field
configurations and equilibrium was ensured by insisting that different
starting configurations yielded consistent results at each temperature.
In the present MC study, we have explored a much larger
antiferromagnetic lattice and modeled our simulations closely after
${\rm Fe}_{1-x}{\rm Zn}_{x}{\rm F}_2$ and the thermal
cycling procedures used in
experiments.  Our large lattice size prevented the averaging over
many random-field configurations, though the results obtained
for a few different configurations yielded consistent results.
We determined the staggered magnetization, $M_s$,
and $C_m$ versus $T$.  The resulting $C_m$ more closely mimics
the symmetric, nearly logarithmic $C_m$ peak of the experiments\cite{sb98}
than the nondivergent cusp found in the earlier MC studies.
In addition, we find $\beta = 0.25\pm 0.03$, in disagreement with
the previous MC studies.

The body-centered-tetragonal magnetic
lattice of ${\rm Fe}_{0.80}{\rm Zn}_{0.20}{\rm F}_2$
is modeled as two cubic sub-lattices delineated as one-dimensional
arrays bit coded to allow for a very large lattice size,
$2L \times L \times L$ with $L = 128$, corresponding to
$3.4 \times 10^{6}$ spins.  Periodic boundary conditions are imposed.
The dilute MC Ising Hamiltonian is
\begin{equation}
H=J_2\sum _{<ij>}\epsilon _i \epsilon _j S_i S_j -h\sum _i \epsilon _i S_i
\end{equation}
where $S_i=\pm 2$ and $\epsilon _i = 1$ if site $i$ is occupied and zero
if not, $J_2=3.17$~K and $h=8.73$~K.  This approximates the
experimental system which has one dominant
Heisenberg exchange interaction, $J_2$, and a large single-ion
anisotropy.  The very small, frustrating interactions,$J_1$ and
$J_3$, are not included. The MC value for $J_2$
is $0.60$ times that of $\rm {FeF_2}$ so that $T_c(0)$ corresponds roughly
to that of the experimental system.  The value of $h$ is chosen
to have roughly the same effect as a field $H=10$~T
in the real system, as determined by the fields at which
low temperature spin flips occur.  This field also yields a shift
$T_c(0)-T_c(H)$ consistent with $H=10$~T in the
real system\cite{fkj91}.  Henceforth we refer to the applied
uniform field as $H=10$~T.

For $H=0$, the sample is cooled to low temperatures in steps of
$0.01$~K while magnetic sites are randomly visited an average
of $N$ times, where $2000<N<5000$,
and flipped with a probability given by the metropolis algorithm.
Similarly, the sample is heated from an ordered state to above
the transition.  No hysteresis is observed at $H=0$ using these
two procedures.  For $H>0$, two thermal cycling procedures are
used to mimic those of the experiments.  Each site is  visited
$5000<N<10000$ times per $0.01$~K temperature step.
For zero-field cooling (ZFC), the sample is started at $H=0$ in a well
ordered state (in experiments this state is prepared by cooling
in $H=0$), the field is subsequently applied,
and the sample is heated through $T_c(H)$.  In field-cooling
(FC), the sample is cooled with $H$ applied.  We observe no
significant hysteresis with these two procedures, consistent
with the experiments.  $M_s$ and $C_m$ are calculated for
each temperature.  The simulations are performed on
Linux/Gnu computers with CPU speeds between 350 to 450 MHz.
Individual runs with $L=128$ at 5000 MC steps per spin and
covering 10K in $T$ typically take one week of CPU time.
We have not seen any significant dependence on $N$ in the range
studied.

Figure 1 shows $C_m$ versus $T$ for $L=128$ and $N=5000$ for $H=0$
and for ZFC and FC at $H=10$~T.  Each point represents a point-by-point derivative
of the energy which is itself averaged over $0.1$~K intervals.
The behavior is very similar to that observed in
birefringence and pulsed heat experiments\cite{sb98}.  Namely, the
$H=0$ behavior exhibits the asymmetric cusp ($\alpha<0$)
characteristic of the random-exchange Ising model and, for $H>0$,
a much more symmetric peak near $T_c(H)$ consistent with the
symmetric, nearly logarithmic divergence.  Previous MC studies
employing finite size scaling did not yield this behavior.

Figure 2 shows the ZFC $M_s$ versus $T-T_{c}$ for $L=128$ and
$N=5000$ for $H=0$ and $H=10$~T.
Fits of the data for various ranges of $|t|$ within $10^{-3}<|t|<10^{-1}$,
yield the critical order-parameter exponent
$\beta = 0.35 \pm 0.04$ for $H=0$ (REIM) and $\beta = 0.25 \pm 0.03$
at $H=10$~T (RFIM).  The $T_c(H)$ values used in the fits were
constrained by the fits to $C_m(H)$

To demonstrate that the results of the simulations are close to
equilibrium, we did a simulation just below the transition
for $L=128$, $H=10$~T, and $T=60.5$.  One lattice was started
fully ordered, another fully ordered but with the
sublattices reversed, and one randomly ordered.
All three lattices converged to the same energy and $M_s$ within
expected fluctuations after $5000$ steps per spin.

Using $\beta = 0.25 \pm 0.03$ from MC
and $\alpha = 0.02 \pm 0.02$ and $\gamma = 1.58 \pm 0.08$ from experiments,
we obtain $2\beta + \gamma +\alpha = 2.08 \pm 0.14$,
which is consistent with the Rushbrooke scaling relation
as an equality.  Further studies will be needed to 
understand why the MC simulations with large antiferromagnetic
lattices and thermal
cycling yield different results than the earlier ferromagnetic MC
simulations with finite scaling analyses\cite{r95}.  For 
simulations of RFIM critical behavior
employing dilute antiferromagnets, the magnetic
concentration should be kept well above the percolation
threshold concentration for vacancies.

This work has been supported by the Department of Energy
Grant No. DE-FG03-87ER45324.

\newpage

\begin{figure}[h]
\centerline{\hbox{
\psfig{figure=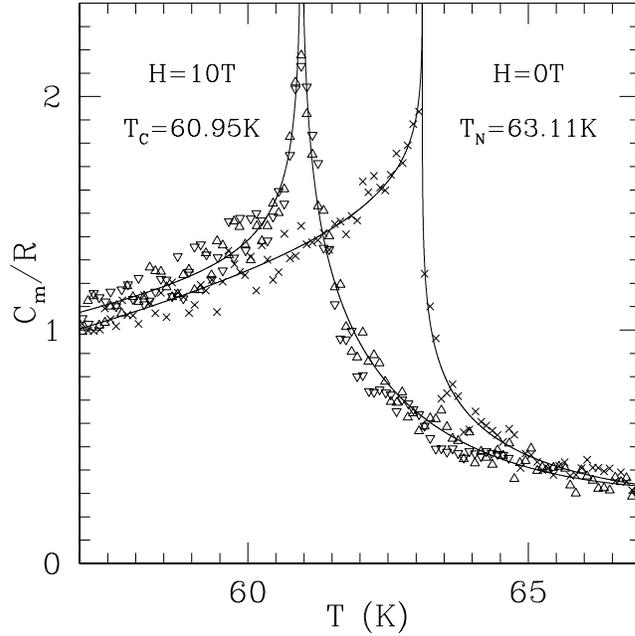,height=3.5in}
}}
\caption{\setlength{\baselineskip}{30pt} $C_m$ vs.\
$T$ for $H=0$ and for ZFC (triangles) and
FC (inverted triangles) at $10$~T with
$L=128$ and $x=0.80$. Note the symmetric peak for $H=10$~T.
The curves are fits to the data with $\alpha = -0.09$ and
$0.00$ for the $H=0$ and $H=10T$ cases, respectively.}
\end{figure} 

\begin{figure}[h]
\centerline{\hbox{
\psfig{figure=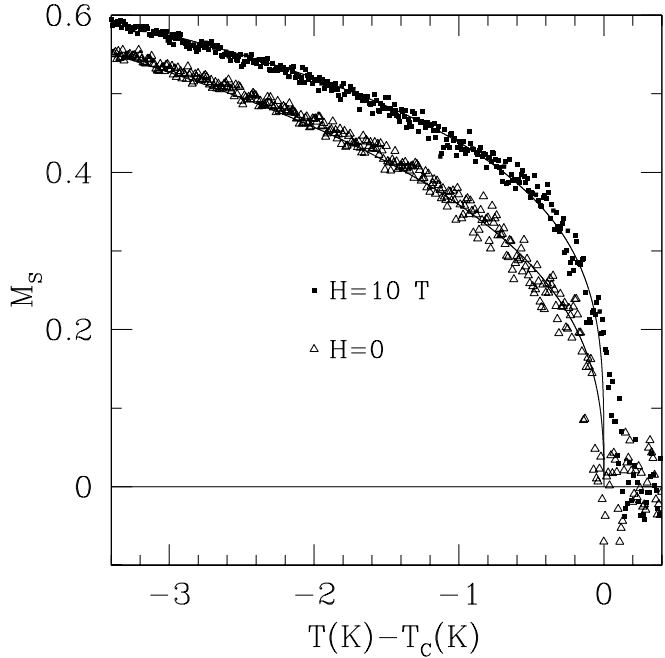,height=3.5in}
}}
\caption{\setlength{\baselineskip}{30pt} $M_s$ vs.\ $T-T_{c}$
for $H=0$ and ZFC at $H=10$~T
with $L=128$ and $x=0.80$.  The curves are fits to the data
with $\beta = 0.35$ and
$0.25$ for the $H=0$ and $H=10T$ cases, respectively.}
\end{figure}

\end{document}